\documentstyle[12pt,epsf]{article} 

\def\lsim{\mathrel{\lower2.5pt\vbox{\lineskip=0pt\baselineskip=0pt 
           \hbox{$<$}\hbox{$\sim$}}}} 
\def\gsim{\mathrel{\lower2.5pt\vbox{\lineskip=0pt\baselineskip=0pt 
           \hbox{$>$}\hbox{$\sim$}}}}

\def\ap{a^{\prime}}

\def\cp{c^{\prime}}

\def\app{a^{\prime\prime}}

\def\cpp{c^{\prime\prime}}

\def\k{\kappa}

\def\d{\delta}
\def\hp{\hat{\psi}}
\def\tp{\tilde{\psi}}
\def\L{\Lambda}

\begin{document} 
\begin{flushright}
DPNU-02-02\\ hep-th/0202166
\end{flushright}

\vspace{10mm}

\begin{center}
{\Large \bf 
 Linearized gravity in flat braneworlds with anisotropic brane tension}

\vspace{20mm}
Masato ITO
\footnote{E-mail address: mito@eken.phys.nagoya-u.ac.jp}
\end{center}

\begin{center}
{
\it 
{}Department of Physics, Nagoya University, Nagoya, 
JAPAN 464-8602 
}
\end{center}

\vspace{25mm}

\begin{abstract}
 We study the four-dimensional gravitational fluctuation
 on anisotropic brane tension embedded in braneworlds with vanishing
 bulk cosmological constant.
 In this setup, warp factors have two types (A and B) and we point out 
 that the two types correspond to positive and negative tension brane,
 respectively.
 We show that volcano potential in the model of type A has singularity
 and the usual Newton's law is reproduced by the existence of
 normalizable zero mode.
 While, in the case of type B, the effective Planck scale is infinite
 so that there is no normalizable zero mode. 
\end{abstract} 
\newpage

 Motivated by the Randall and Sundrum model 
 \cite{Randall:1999ee,Randall:1999vf},
 the localization of the four-dimensional gravity is widely investigated
 in the framework of warped  braneworlds.
 In the original Randall-Sundrum braneworld of $AdS_{5}$ with an
 infinite extra dimension, massless gravity is trapped on the brane and
 massive gravity has continuous spectrum, consequently, the usual
 four-dimensional Newton's law can be seen by observer at large distance
 scale.
 Following remarkable development of $AdS/CFT$ correspondence
 \cite{Maldacena:1997re,Witten:1998qj,Gubser:1998bc} that
 the gravitational theories in $AdS$ space are dual to conformal field
 theories on the boundary, the Randall-Sundrum model becomes attractive
 model in particle physics and cosmology. 
 Moreover there have been many proposals to show how four-dimensional
 gravity emerges in warped braneworld with infinite extra dimensions
 \cite{Lykken:1999nb,Giddings:2000mu,Csaki:2000fc,Ito:2001nc}.
 For example, the cases of $dS_{4}$ brane and $AdS_{4}$ brane embedded
 in $AdS_{5}$ are investigated in detail
 \cite{Karch:2000ct}.
 In the case of $dS_{4}$ brane, a massless bound state gravity exists.
 While in the case of $AdS_{4}$ brane, the massless bound state gravity
 don't exist and massive gravity has discrete spectrum because the
 volcano potential blows up at the boundary as if box-like potential. 
 Thus the behavior of gravity is governed by the volcano potential in
 Schr$\ddot{\rm o}$dinger equation for gravitational fluctuation.

 In this letter, we consider the braneworlds with vanishing
 cosmological constant in contrast with the Randall-Sundrum model
 with negative bulk cosmological contrast. 
 We study the fluctuation of gravity on the brane with anisotropic brane
 tension embedded in this braneworld, and show whether the usual
 Newton's law is reproduced or not by investigating the volcano potential of
 this setup.

 We consider a single $(3+n)$-brane embedded in the $(5+n)$-dimensional world.
 The brane is located at  $y=0$, where the $y$-direction is assumed to
 have $Z_{2}$ symmetry.
 Here the ansatz for $(5+n)$-dimensional metric is written in the
 following form \cite{Ito:2001fd}
 \begin{eqnarray}
  ds^{2}&=&a^{2}(y)\eta_{\mu\nu}dx^{\mu}dx^{\nu}
        +c^{2}(y)\sum^{n}_{i=1}\;dz^{2}_{i}+dy^{2}\nonumber\\
  &\equiv&g_{MN}dx^{M}dx^{N}\label{eqn1}\,.
 \end{eqnarray}
 Note that the usual four-dimensional spacetime preserves 
 Poinc$\acute{\rm a}$re invariance and the warp factor of
 four-dimensional spacetime is
 different from that of the extra $n$ dimensional space.
 It is assumed that $z$-directions are compactified in the same radius
 $R\sim M^{-1}_{\rm pl}$, where $M_{\rm pl}$ is Planck scale.
 We give Einstein equation of this setup,
 ${\cal R}_{MN}-\frac{1}{2}g_{MN}{\cal R}=-\k^{2}
  \L g_{MN}-\k^{2}T_{MN}$, where $\L$ is the bulk
 cosmological constant and $T_{MN}$ is the energy-momentum tensor of
 the brane, and $1/\kappa^{2}$ is the higher dimensional gravitational
 constant which has mass dimension $3+n$.
 Here it is assumed that the distribution of the brane tension of the
 brane with $(4+n)$-dimensional spacetime is anisotropic and that the
 effects of matters on the brane are neglected \cite{Ito:2001fd}.
 Hence the energy-momentum tensor of brane is given by
 \begin{eqnarray}
  T^{M}_{N}=\d(y)\;
  diag(\;V,V,V,V,\underbrace{V^{\ast},\cdots,V^{\ast}}_{n},0\;)\,,
  \label{eqn2}
 \end{eqnarray}
 where $V$ and $V^{\ast}$ represent the brane tension of
 four-dimensional spacetime and the one of extra $n$-dimensional space,
 respectively.

 For $\L=0$, solving Einstein equation with metric of Eq.(\ref{eqn1}),
 warp factors have two types as follows \cite{Ito:2001fd}
 \begin{eqnarray}
  {\rm Type\;A}\hspace{1cm}
  a(y)=\left(\;1-\frac{|y|}{d}\;\right)^{k}\,,\hspace{1cm}
  c(y)=\left(\;1-\frac{|y|}{d}\;\right)^{l}\,,\label{eqn3}
 \end{eqnarray}
 and
 \begin{eqnarray}
 {\rm Type\;B}\hspace{1cm}
  a(y)=\left(\;1+\frac{|y|}{d}\;\right)^{k}\,,\hspace{1cm}
  c(y)=\left(\;1+\frac{|y|}{d}\;\right)^{l}\,,\label{eqn4}
 \end{eqnarray}
 where $d$ is the positive constant and the warp factors are normalized
 to be unity at $y=0$. 
 For both types,  $k$ and $l$ must be determined by the following
 equations
 \begin{eqnarray}
  4k+nl=1\,,\hspace{1cm}4k^{2}+nl^{2}=1\,.\label{eqn5}
 \end{eqnarray}
 Note that the equations for the exponents are closely
 similar to the Kasner solutions appearing in the higher dimensional
 cosmology.
 The jump condition at $y=0$ due to the delta-function leads to
 the relation between brane tensions as follows
 \begin{eqnarray}
  \frac{V}{1-k}=\frac{V^{\ast}}{1-l}
  =\pm\frac{2}{\k^{2}d}\,,\label{eqn6}
 \end{eqnarray}
 where $+$($-$) corresponds to the type A(B).
 From the above equation, due to positive constant $d$,
 type A and type B correspond to positive tension branes and negative
 tension branes, respectively.
 This situation is similar to the Randall-Sundrum model with single
 brane because $-(+)$ in warp factor $e^{\mp 2|y|/\rho}$ obtained in
 the model corresponds to positive (negative) tension brane,
 where $\rho$ is the radius of $AdS$.   
 From Eq.(\ref{eqn6}), anisotropic brane tensions are related through the
 number of $n$.
 Solving Eq.(\ref{eqn5}), we have 
 \begin{eqnarray}
  k=k_{1,2}=\frac{2\pm\sqrt{n(n+3)}}{2(n+4)}\,,\hspace{1cm}
  l=l_{1,2}=\frac{n\mp 2\sqrt{n(n+3)}}{n(n+4)}\,,\label{eqn7}
 \end{eqnarray}
 where the subscript $1$ and $2$ correspond to the choice of taking
 upper sign and lower sign, respectively.
 Thus the forms of warp factors depend on both $n$ and the choice of
 sign.
 From Eq.(\ref{eqn7}), for arbitrary positive integer $n\geq 1$,
 always $k_{1,2}<1$ and $l_{1,2}<1$.
 
 To see the four-dimensional gravity on the brane, the fluctuation
 around the background metric is given by
 $a^{2}(y)\eta_{\mu\nu}+h_{\mu\nu}(x,y)$.
 Here the dependence of $z$-directions in $h_{\mu\nu}$ are neglected.
 Although the gravity in the bulk has massive KK-modes along $n$
 compactified $z$-directions with Planck size, these dimensions are
 invisible at low energy.
 Namely, at large distance, it is considered that the fluctuations of
 $z$-directions are negligible at this stage. 
 Later we discuss this point.

\begin{figure}
      \epsfxsize=6cm
\centerline{\epsfbox{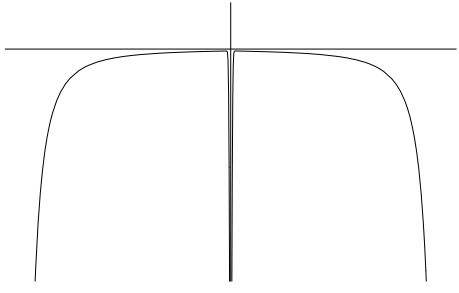}}
\caption{Volcano potential of type A}
\label{fig1} 
\end{figure}

\begin{figure}
      \epsfxsize=6cm
\centerline{\epsfbox{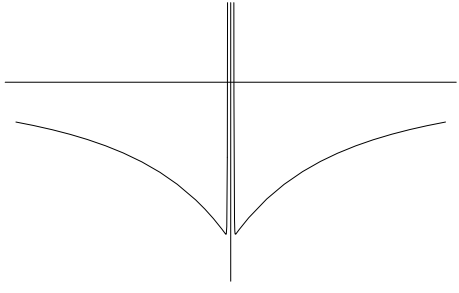}}
\caption{Volcano potential of type B}
\label{fig2} 
\end{figure}

 Setting $h_{\mu\nu}(x,y)=h_{\mu\nu}(x)c^{-n/2}(y)\hp(y)$,
 the linearized equation of motion for transverse-traceless mode
 is given by
 \begin{eqnarray}
  \frac{d^{2}}{dy^{2}}\hp (y)
  +\left[\;\frac{m^{2}}{a^{2}}-8\frac{\app}{a}
  +4\left(\frac{\ap}{a}\right)^{2}
  -\frac{5}{2}n\frac{\cpp}{c}
  -\frac{1}{4}n(n-2)\left(\frac{\cp}{c}\right)^{2}\right.\nonumber\\
  \left.
  -2\k^{2}V\d(y)\frac{}{}\right]\hp (y)=0\,.\label{eqn8}
 \end{eqnarray}
 Performing the change of variables, $\hp(y)=\tp(u)(1\mp|y|/d)^{k/2}$ and
 $(1\mp|y|/d)^{1-k}=1\mp(1-k)|u|/d$, we can obtain familiar non-relativistic
 quantum mechanics problem as follows,
 \begin{eqnarray}
  \left[\;-\frac{d^{2}}{du^{2}}+V(u)\;\right]\tp (u)
 =m^{2}\tp (u)\,,
  \label{eqn9}
 \end{eqnarray}
 where $m^{2}$ is the four-dimensional mass.
 The potential $V(u)$ in Schr$\ddot{\rm o}$dinger equation is so-called
 the volcano potential which can be written in terms of $k$ using
 Eq.(\ref{eqn5})
 \begin{eqnarray}
  V(u)=-\frac{1}
  {\displaystyle 4\left(\frac{d}{1-k}\mp|u|\right)^{2}}
       \mp\frac{1-k}{d}\d(u)\,.\label{eqn10}
 \end{eqnarray}
 Here upper sign (lower sign) corresponds to the type A(B)
 as shown in Figure.\ref{fig1}(\ref{fig2}).
 In the volcano potential of both types, the eigenvalue becomes positive
 definite value, $m^{2}\geq 0$.
 This is because the bulk equation of Eq.(\ref{eqn9}) is expressed as 
 $Q^{\dagger}Q\tilde{\psi}=m^{2}\tilde{\psi}$, where
 $Q=d/du\pm 2^{-1}(d/(1-k)\mp u)^{-1}$.
%
%
%
%
 Thus since there are no tachyon modes, the gravitational background
 becomes stable.
 Due to $k<1$ for positive integer $n$, the delta function of
 type A becomes potential of attractive force via positive tension
 brane and the delta function of type B is potential of repulsive force
 via negative tension brane.
 In the case of type A shown in Figure.\ref{fig1},
 the $u$ coordinate runs from $-d/(1-k)$ to
 $d/(1-k)$, both side of the volcano potential goes negative infinity at
 $u=\pm d/(1-k)$.
 If we take the upper sign in Eq.(\ref{eqn7}), $k_{1}>0$ and $l_{1}<0$,
 then $c\rightarrow\infty$ at $y=\pm d$ ( this implies that $u=\pm
 d/(1-k)$ ). 
 On the other hand, taking the lower sign, $k_{2}<0$ and $l_{2}>0$,
 $a\rightarrow\infty$ at the point.
 This implies that the singularity at $y=d$ becomes curvature singularity.
 However, in this case of $n=1$, $k_{2}=0$ and $l_{2}=1$, the warped metrics
 have no singularity.
 In the case of type B shown in Figure.\ref{fig2},
 the $u$ coordinate runs from $-\infty$ to
 $+\infty$, $|V(u)|\rightarrow 0$ for $|u|\rightarrow \infty$.
 For arbitrary $n$, the warped metrics have no singularity.

 We solve the Schr$\ddot{\rm o}$dinger equation of linearized gravity
 with volcano potential for type A and B. 
 For zero mode wavefunction, we obtain
 \begin{eqnarray}
  \tp_{0}(u)\propto\left(\frac{d}{1-k}\mp|u|\right)^{\frac{1}{2}}
  \label{eqn11}\,.
 \end{eqnarray}
 where the jump condition of delta function is imposed.
 The wavefunction of the type A (upper sign) is normalizable over
 limited range $|u|\leq d/(1-k)$. 
 On the other hand, the wavefunction of the type B (lower sign) is
 non-normalizable because of $\tp_{0}(u)\rightarrow\infty$ for
 $|u|\rightarrow\infty$.
 Namely, in the case of type B, the normalizable zero mode wavefunction
 cannot exist.
 In the case of type A, the normalizable zero mode wavefunction is given
 by
 \begin{eqnarray}
  \tp_{0}(u)=\frac{1-k}{d}\left(\frac{d}{1-k}-|u|\right)^{\frac{1}{2}}
  \label{eqn12}\,.
 \end{eqnarray}
 For massive mode wavefunction $m^{2}>0$, from Eqs.(\ref{eqn9}) and
 (\ref{eqn10}), the wavefunction of the gravity is written in terms of
 the superposition of Bessel functions as follows,
 \begin{eqnarray}
  \tp^{\mp}_{m}(u)&=&
  N^{\mp}_{m}\left(\frac{d}{1-k}\mp|u|\right)^{1/2}\nonumber\\
  &&\times
  \left\{-\frac{Y_{1}\left(\displaystyle\frac{md}{1-k}\right)}
  {J_{1}\left(\displaystyle\frac{md}{1-k}\right)}
  J_{0}\left(m\left(\frac{d}{1-k}\mp|u|\right)\right)
  +Y_{0}\left(m\left(\frac{d}{1-k}\mp|u|\right)\right)
  \right\}\,,  \nonumber\\
 &&
  \label{eqn13}
 \end{eqnarray}
 where upper sign (lower sign) corresponds to the type A(B) and
 $N^{\mp}_{m}$ is normalization factor, and the jump condition of
 delta function is imposed.
 Thus massive modes are continuous, namely, these aren't bound states. 
 From Eq.(\ref{eqn13}), the normalization factor of type A is expressed as
 \begin{eqnarray}
 N^{-}_{m}=\left[2\int^{u_{0}}_{0}dt\;
           t\left\{-\frac{Y_{1}(mu_{0})}{J_{1}(mu_{0})}
            J_{0}(t)+Y_{0}(t)\right\}^{2}
 \right]^{-1/2}\,,\label{eqn14}
 \end{eqnarray}
 where $u_{0}=d/(1-k)$.
 The normalization factor of type B is given by
 \begin{eqnarray}
  N^{+}_{m}=\frac{\sqrt{m}\;J_{1}(mu_{0})}
  {\sqrt{J^{2}_{1}(mu_{0})+Y^{2}_{1}(mu_{0})}}\,.\label{eqn15}
 \end{eqnarray}
 The difference between the normalization factor of type A and B
 comes from the range of $u$-coordinate.

 We are interested in the realization of the four-dimensional Newton's law
 in the framework of model considered here.
 In the type A, the gravitational potential via zero mode is
 given by $V\propto G_{N}/r$ at large distance scale, where
 $G_{N}\sim \kappa^{2}|\tilde{\psi}_{0}(0)|^{2}/R^n
 =\kappa^{2}(1-k)/(R^{n}d)$. 
 The contribution of massive mode gives small correction to the
 potential, it is given by
 $\int^{\infty}_{0}dm\;|\tilde{\psi}^{-}_{m}(0)|^{2}e^{-mr}/r$.  
 In the type B, since there is no normalizable zero mode,
 the effective four-dimensional Planck scale is infinite.
 Thus the four-dimensional Newton's law cannot be reproduced.

 As mentioned previously, the $z$-dependence of wavefunction
 $h_{\mu\nu}(x,y)$ is neglected here.
 If the $z$-dependence is imposed, the wavefunctions
 of $z$-directions $\varphi(z)$ become plane waves with KK-mode
 expansion, namely, $\varphi(z)\sim \sum_{p}e^{i p_{j}z^{j}/R}$,
 where $p_{j}\in {\rm Z}$ and $j=1,\cdots ,n$. 
 Furthermore, the eigenvalue of Eq.(\ref{eqn9})
 must be replaced in $m^{2}\rightarrow m^{2}+\sum |\vec{p}|^{2}/R^{2}$,
 thus the eigenvalues have both continuous modes and discrete KK-modes.
 Consequently, the volcano potential of Eq.(\ref{eqn10}) is not modified even
 if the fluctuations of $z$-directions are taken into account.
 Although the corrections to Newton's law via massive modes 
 have the contributions of discrete KK-modes, the contributions are
 sufficiently suppressed at large distance $r\gg M^{-1}_{\rm pl}$.

 In conclusion, we considered a single $(3+n)-$brane embedded in
 $(5+n)$-dimensional world with vanishing cosmological constant.
 It assumed that the brane has anisotropic brane tension and that the
 warp factor of four-dimensional spacetime $a(y)$ is different from one
 of $n$-dimensional spaces $c(y)$.
 In this setup, warp factors have two types (A and B) and we pointed out
 that the type A and the type B correspond to the positive tension
 brane and the negative tension brane, respectively.
 The volcano potential of the type A has singularity, $y$-coordinate is
 effectively truncated.
 At this stage we don't have quantitative to say about the physical
 implications of this singularity, we will describe it elsewhere.
 Investigating the linearized gravity around background metric,
 there is a normalizable zero mode wavefunction, consequently,
 the usual four-dimensional gravity is reproduced. 
 On the other hand, the volcano potential of type B has no singularity, 
 and it is shown that there is no normalizable zero mode.
 Since the effective four-dimensional Planck scale is infinite, the
 usual four-dimensional gravity cannot be reproduced.
 In contrast with Randall-Sundrum model,
 we proposed a simple model in which the localization of gravity is
 realized in braneworld with not $\L<0$ but $\L=0$, where $\L$ is 
 the bulk cosmological constant. 
 Thus the setup with warped geometry is completely different from 
 the setup with non-warped geometry 
 Finally we describe a comment.
 Since the setup considered here doesn't include the fluctuations of gravity
 corresponding to all coordinates, four-dimensional tensor structure
 cannot be explicitly reproduced.
 Furthermore these fluctuations play an important role in brane
 stabilization.
 In the future this point will be investigated elsewhere.

%

\end{document}